\documentclass[%
aps, 
superscriptaddress,
 sd,%
 amsmath,amssymb,
 reprint,%
twocolumn, 
twoside,
author-numerical,%
]{revtex4-1}

\usepackage{graphicx}
\usepackage[colorlinks = true,
            linkcolor = blue,
            urlcolor  = blue,
            citecolor = blue,
            anchorcolor = blue]{hyperref}

\newcommand{\be}{\begin{equation}}
\newcommand{\ee}{\end{equation}}

\begin{document}
\newcommand{\transition}{${}^5D_0-{}^7F_0{\,}$}
\newcommand{\transitionstrong}{${}^5D_0-{}^7F_2{\,}$}
\newcommand{\eucrystal}{Eu\textsuperscript{3+}:Y\textsubscript{2}O\textsubscript{3}{ }}
\newcommand{\eu}{Eu\textsuperscript{3+}}
\newcommand{\ercrystal}{Er\textsuperscript{3+}:Y\textsubscript{2}O\textsubscript{3}{ }}


\title{Cavity-enhanced spectroscopy of a few-ion ensemble in Eu\textsuperscript{3+}:Y\textsubscript{2}O\textsubscript{3}} 



\author{Bernardo Casabone}
\affiliation{ICFO-The Institute of Photonic Sciences, Mediterranean Technology park, 08860 Castelldefels, Spain}
\affiliation{Max-Planck-Institut f{\"u}r Quantenoptik, Hans-Kopfermann-Stra{\ss}e 1, 85748 Garching, Germany}

\author{Julia Benedikter}
\affiliation{Fakult{\"a}t f{\"u}r Physik, Ludwig-Maximilians-Universit{\"a}t, Schellingstra{\ss}e 4, 80799 M{\"u}nchen, Germany}
\affiliation{Max-Planck-Institut f{\"u}r Quantenoptik, Hans-Kopfermann-Stra{\ss}e 1, 85748 Garching, Germany}
\affiliation{Karlsruher Institut f{\"u}r Technologie, Physikalisches Institut, Wolfgang-Gaede-Str. 1, 76131 Karlsruhe, Germany}

\author{Thomas H{\"u}mmer}
\affiliation{Fakult{\"a}t f{\"u}r Physik, Ludwig-Maximilians-Universit{\"a}t, Schellingstra{\ss}e 4, 80799 M{\"u}nchen, Germany}
\affiliation{Max-Planck-Institut f{\"u}r Quantenoptik, Hans-Kopfermann-Stra{\ss}e 1, 85748 Garching, Germany}

\author{Franziska Oehl}
\affiliation{Fakult{\"a}t f{\"u}r Physik, Ludwig-Maximilians-Universit{\"a}t, Schellingstra{\ss}e 4, 80799 M{\"u}nchen, Germany}

\author{Karmel de Oliveira Lima}
\affiliation{Universit{\'e} PSL, Chimie ParisTech, CNRS, Institut de Recherche de Chimie Paris, 75005 Paris, France}

\author{Theodor W. H{\"a}nsch}
\affiliation{Fakult{\"a}t f{\"u}r Physik, Ludwig-Maximilians-Universit{\"a}t, Schellingstra{\ss}e 4, 80799 M{\"u}nchen, Germany}
\affiliation{Max-Planck-Institut f{\"u}r Quantenoptik, Hans-Kopfermann-Stra{\ss}e 1, 85748 Garching, Germany}

\author{Alban Ferrier}
\affiliation{Universit{\'e} PSL, Chimie ParisTech, CNRS, Institut de Recherche de Chimie Paris, 75005 Paris, France}
\affiliation{Sorbonne Universit{\'e}, 75005 Paris, France}

\author{Philippe Goldner}
\affiliation{Universit{\'e} PSL, Chimie ParisTech, CNRS, Institut de Recherche de Chimie Paris, 75005 Paris, France}
\affiliation{Sorbonne Universit{\'e}, 75005 Paris, France}

\author{Hugues de Riedmatten}
\affiliation{ICFO-The Institute of Photonic Sciences, Mediterranean Technology park, 08860 Castelldefels, Spain}
\affiliation{ICREA-Instituci{\'o} Catalana de Recerca i Estudis Ava\c{c}ats, 08015 Barcelona, Spain}

\author{David Hunger}
\affiliation{Karlsruher Institut f{\"u}r Technologie, Physikalisches Institut, Wolfgang-Gaede-Str. 1, 76131 Karlsruhe, Germany}
\email{david.hunger@kit.edu}

\date{\today}

\begin{abstract}
We report on the coupling of the emission from a single europium-doped nanocrystal to a fiber-based microcavity under cryogenic conditions. 
As a first step, we study the properties of nanocrystals that are relevant for cavity experiments and show that embedding them in a dielectric thin film can significantly reduce scattering loss and increase the light matter coupling strength for dopant ions. The latter is supported by the observation of a fluorescence lifetime reduction, which is explained by an increased local field strength. 
We then couple an isolated nanocrystal to an optical microcavity, determine its size and ion number, and perform cavity-enhanced spectroscopy by resonantly coupling a cavity mode to a selected transition.
We measure the inhomogeneous linewidth of the coherent \transition transition and find a value that agrees with the linewidth in bulk crystals, evidencing a high crystal quality. We detect the fluorescence from an ensemble of few ions in the regime of power broadening and observe an increased fluorescence rate consistent with Purcell enhancement. 
The results represent an important step towards the efficient readout of single rare-earth ions with excellent optical and spin coherence properties, which is promising for applications in quantum communication and distributed quantum computation.
\end{abstract}


\maketitle 


\section{Introduction}
Rare earth ion-doped crystals constitute a promising solid state system for quantum information processing and networking 
\cite{Goldner20151}.
They provide nuclear spin states with very long coherence times, up to six hours \cite{zhong2015}, in which quantum information can be stored\cite{kolesov2013, laplane2015, gundougan2015}. 
They also offer $4f-4f$ optical transitions with exceptionally good coherence properties, up to $4.3$~ms \cite{bottger2009}, that can be used as a photonic interface for the spin states. 
Furthermore, electric dipole interactions can be used for an excitation blockade similar to the one studied for Rydberg atoms, and are thus a resource to realize quantum gates \cite{ohlsson2002, longdell2004}.
However, due to the dipole-forbidden transitions, the excited-state lifetime is long, leading to very low emission rates. This has limited most experiments to macroscopic ensembles so far,
which hinders scalability towards multi-qubit systems. 


Recently, individual rare earth ions have been detected either by upconversion spectroscopy \cite{kolesov2012}, direct high-resolution spectroscopy of $4f-4f$ transitions \cite{utikal2014,nakamura2014}, or by using strong $5d-4f$ transitions \cite{kolesov2013,siyushev2014}. For the realization of spin-photon interfaces, the coherent $4f-4f$ transitions appear to be best suited. 
However, their low scattering rate makes single ion detection in free space challenging. 
Moreover, the excited state lifetime is typically significantly longer than the coherence time, deteriorating the indistinguishability of the emitted photons.
Both issues can be overcome by coupling the emitters to optical microcavities. 
This increases the low scattering rates to achieve practically useful detection rates \cite{zhong2015, dibos2017, zhong2018}.
Coupling the emitters to optical microcavities increases the spontaneous emission rate by the Purcell factor 
$C = \zeta \frac{3\lambda^3}{4\pi^2}\frac{Q}{V_m} $,
where $\lambda$ is the emission wavelength, $Q$ the quality factor of the resonator, $V_m$ its mode volume, and $\zeta$ the branching ratio of the respective transition. 
Additionally, the emission is coupled to a well-collectible cavity mode, yielding a collection efficiency given by $\beta = \eta C/(C+1)$ with the cavity outcoupling efficiency $\eta$. Near-unity collection efficiency can be reached for sufficiently large $C$ and $\eta$. Finally, for $C > 2\gamma_h\tau$, where $\tau $ is the excited state lifetime and $\gamma_h$ is the homogeneous linewidth, emitted photons are Fourier limited~\cite{zhong2018}, 
thus allowing for interference between photons from different ions as required for scalable entanglement distribution using entanglement swapping~\cite{zukowski1993} as already shown in other systems~ \cite{pan1998, olmschenk2009}.

Here, we report on a promising approach for achieving efficient access to individual or small ensembles of ions by coupling ion-doped nanocrystals to a high-finesse fiber-based Fabry-Perot microcavity \cite{hunger2010, benedikter2017}. 
Fiber cavities can achieve high Purcell factors up to $10^4$, provide open access to the cavity mode for optimal overlap between the ions and the cavity field, and offer full tunability of the resonance frequency to target all regions of the inhomogeneous line. 
We show that introducing small enough nanoparticles into the cavity does not significantly affect the cavity, and allows one to independently optimize the cavity and the sample. In particular, we demonstrate that embedding nanoparticles into a dielectric layer on a cavity mirror is beneficial for both reducing scattering loss and increasing the local field strength at the ions, significantly improving ion-cavity coupling.
Nanoparticles naturally select a small ensemble of ions, such that due to a large ratio between the inhomogeneous and the homogeneous linewidth, individual ions can be addressed via their transition frequency. 
We study \eucrystal nanocrystals, which to date have shown the best optical and spin coherence within a nanoscale host, with a homogeneous linewidth $\gamma_h=45$~kHz \cite{perrot2013,bartholomew2017} and $8$~ms nuclear spin coherence \cite{serrano2017} observed in nanoparticle powders.
Europium also offers a hyperfine structure that can be used to store and process quantum information \cite{lauritzen2012}.
We demonstrate cavity-enhanced low-temperature spectroscopy of a few-ion ensemble in a crystal where the size and thus the contained ion number is known. We estimate the achieved cavity enhancement and deduce the number of ions that contribute to the signal. The approach is promising for efficient single ion detection and could be a basis to realize quantum nodes for quantum communication and information processing.

\begin{figure*}
	\includegraphics[width=1\textwidth]{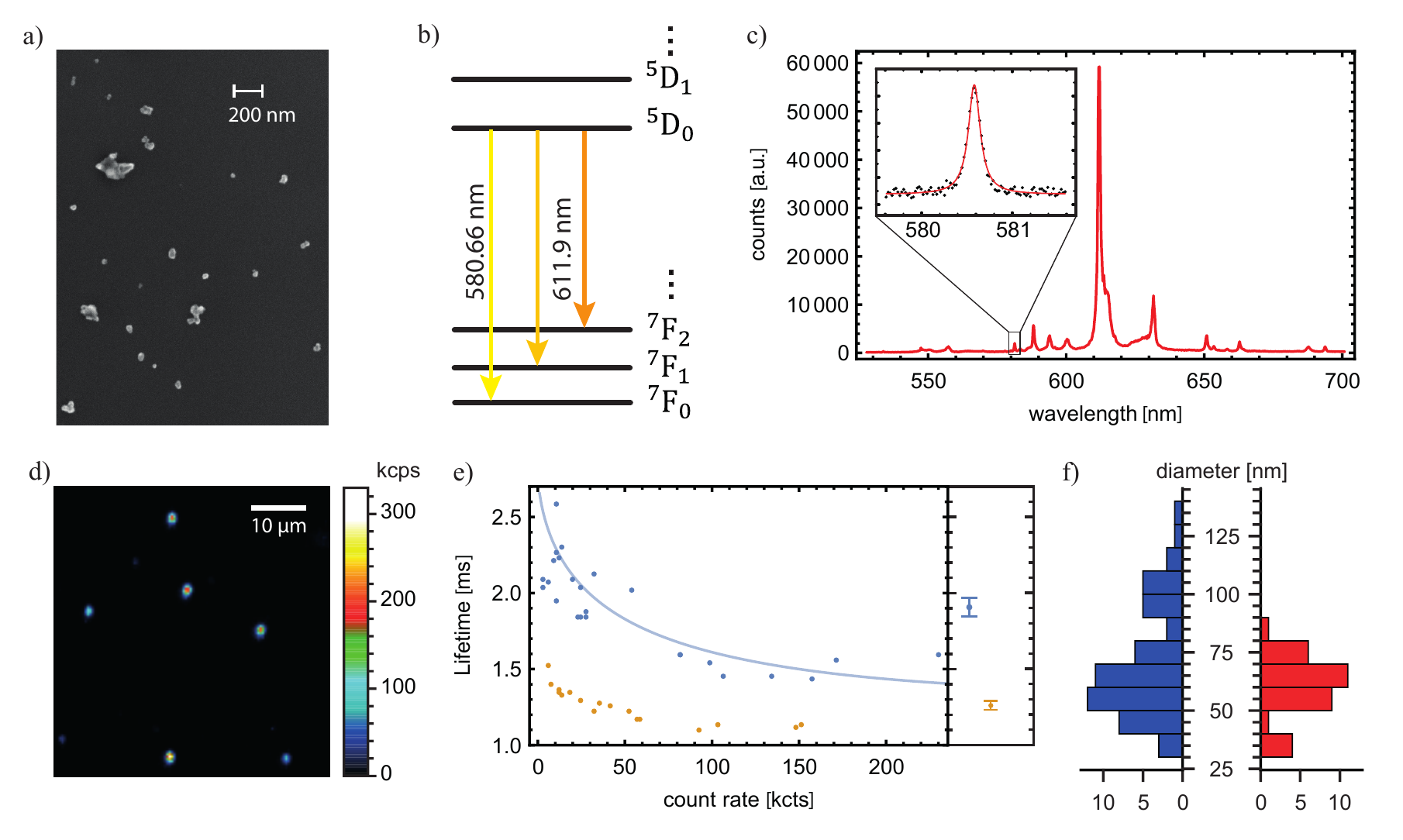}
	\caption{
		{
			a) SEM image of isolated nanocrystals dispersed on a substrate.
			b) Relevant level diagram of \eucrystal, showing the coherent \transition transition at $580$~nm and the strongest \transitionstrong transition at $611$~nm.
			c) Broadband fluorescence spectrum of a single nanocrystal recorded by confocal microscopy. The inset shows the \transition transition in high resolution together with a Voigt profile fit. Ions are off-resonantly excited using a laser at $532$~nm. 
			d) Confocal fluorescence image of individual nanocrystals spin-coated on a mirror. 
			e) Fluorescence lifetime measurements as function of the photon count-rate. 
			Each data point corresponds to a single measurement carried out on a randomly chosen crystal.
			Blue: crystals on silica slip. The blue solid line is a fit to a model. Orange: crystals embedded in a PMMA thin film on a mirror. 
			The right panel shows the average lifetime and standard deviation for both cases. 
			f) Histogram of measured crystal sizes (right panel, red bars) and histogram of calculated crystal sizes from a measured fluorescence distribution.
		}
		\label{fig1}}
\end{figure*}

\section{Single nanocrystal fluorescence, scattering and lifetime}
We investigate the properties of nanocrystals with a confocal setup (air objective with NA=0.9) at room temperature to obtain reference values of optical properties and to optimize the integration of crystals into a microcavity with the aim to minimize scattering loss and maximize the light-matter coupling.

We study $0.5\%$ \eucrystal nanoparticles with an average size of $60$~nm synthesized by homogeneous precipitation \cite{de2015}. Figure~\ref{fig1}(a) shows a scanning electron micrograph of nanocrystals dispersed on a substrate. Figure~\ref{fig1}(d) shows a confocal fluorescence image of individual nanocrystals spin-coated on a mirror. We excite under saturation conditions ($\sim 1$~mW, NA$=0.9$) with $532$~nm light, which is close to the $^5D_1 - ^7F_1$ transition. The emission occurs from ${}^5D_0$ with several transitions into different spin-orbit ground state levels. Figure ~\ref{fig1}(b) shows a simplified scheme of the relevant levels and transitions studied here.
We record spectra of individual crystals and assess the room-temperature linewidth of the \transition transition, which gives a first impression of the crystal quality. 
Figure~\ref{fig1}(c) shows a broadband fluorescence spectrum of a single crystal, and the inset shows the \transition transition in high resolution. 
Even for nanocrystals with very low count rate, which are correspondingly small and thus compatible with cavity experiments, we observe a linewidth of $80$~GHz, which is as narrow as in bulk samples. 

Next, we study a large number of isolated nanocrystals and infer the peak count rate per crystal. 
By looking at the count rate distribution and comparing with a size distribution obtained from SEM images such as the one shown in figure~\ref{fig1}(a), we can get an order of magnitude estimate for the typical count rate per single ion. Figure~\ref{fig1}(f) shows a histogram of the measured crystal size (red, right) and the expected crystal size distribution calculated from the measured fluorescence values (blue, left). Here we assume that the count rate $R$ and crystal radius $r$ are directly related via $R=N R_0 = (4\pi/3) n_0 r^3 R_0$, where $R_0$ is the single ion count rate and $N$ the number of ions. From the size and doping concentration, and knowing that the Y concentration in the C2 site is $n_0=1.6 \times 10^{22}/cm^3$, single crystals of e.g.\, 60~nm diameter contain $N\sim 10^4$ ions. The two distributions match if we assume an excited state population of $20\%$ (several ground state levels are thermally populated, but only one is excited at about one saturation intensity) and an overall fluorescence detection efficiency of $1.5\%$. This value is smaller than the expected maximally achievable detection efficiency of $20\%$ for our setup, which we calculate from the modified dipole emission pattern on a Bragg mirror averaged over dipole orientations, the angular collection efficiency of the microscope objective, the maximal transmission through the optics, and the detector quantum efficiency \cite{benedikter2017}. The smaller experimental value is ascribed to suboptimal collection and transmission efficiency and a potential overestimation of the excited state population. The analysis shows that we detect about 2 photons per second per ion, which at best could be increased to $~100$ photons/s, rendering single ion experiments unpractical under free space conditions.

Two crucial aspects for the cavity experiment are the scattering loss introduced by the nanocrystal and the achievable ion-photon coupling strength, where the latter depends on the partially suppressed local electric field inside the crystal. Indeed, scattering loss and local field suppression both arise from the boundary condition introduced by the nanocrystal surface, and are related to modes outside or inside the crystal, respectively.
In the Rayleigh regime $r\ll\lambda/2\pi$, the scattering loss 
\begin{equation}
B=4\sigma/\pi w_0^2
\end{equation}
can be calculated from the cavity mode waist $w_0$ and the scattering cross section
\begin{equation}
\sigma=\left(\frac{2\pi}{\lambda}\right)^4\frac{\alpha^2}{6\pi},
\end{equation}
with the polarizability of the crystal \cite{wind1987}
\begin{equation}
\alpha=3\epsilon_0 V \frac{n^2-n^2_m}{n^2+2n^2_m}
\end{equation}
with $n$ and $n_m$ the refractive indices of the crystal and surrounding medium, $\epsilon_0$ the vacuum dielectric constant, and $V=4/3\pi r^3$ the volume of the crystal.
It is apparent that particles embedded in a dielectric layer lead to a reduced scattering loss. For the case of Y$_2$O$_3$ particles ($n=1.93$) embedded in a PMMA layer ($n_m=1.49$), the calculation above yields a reduction of $B$ by a factor of 6.7 compared to a crystal in air.

\begin{figure*}
	\includegraphics[width=\textwidth]{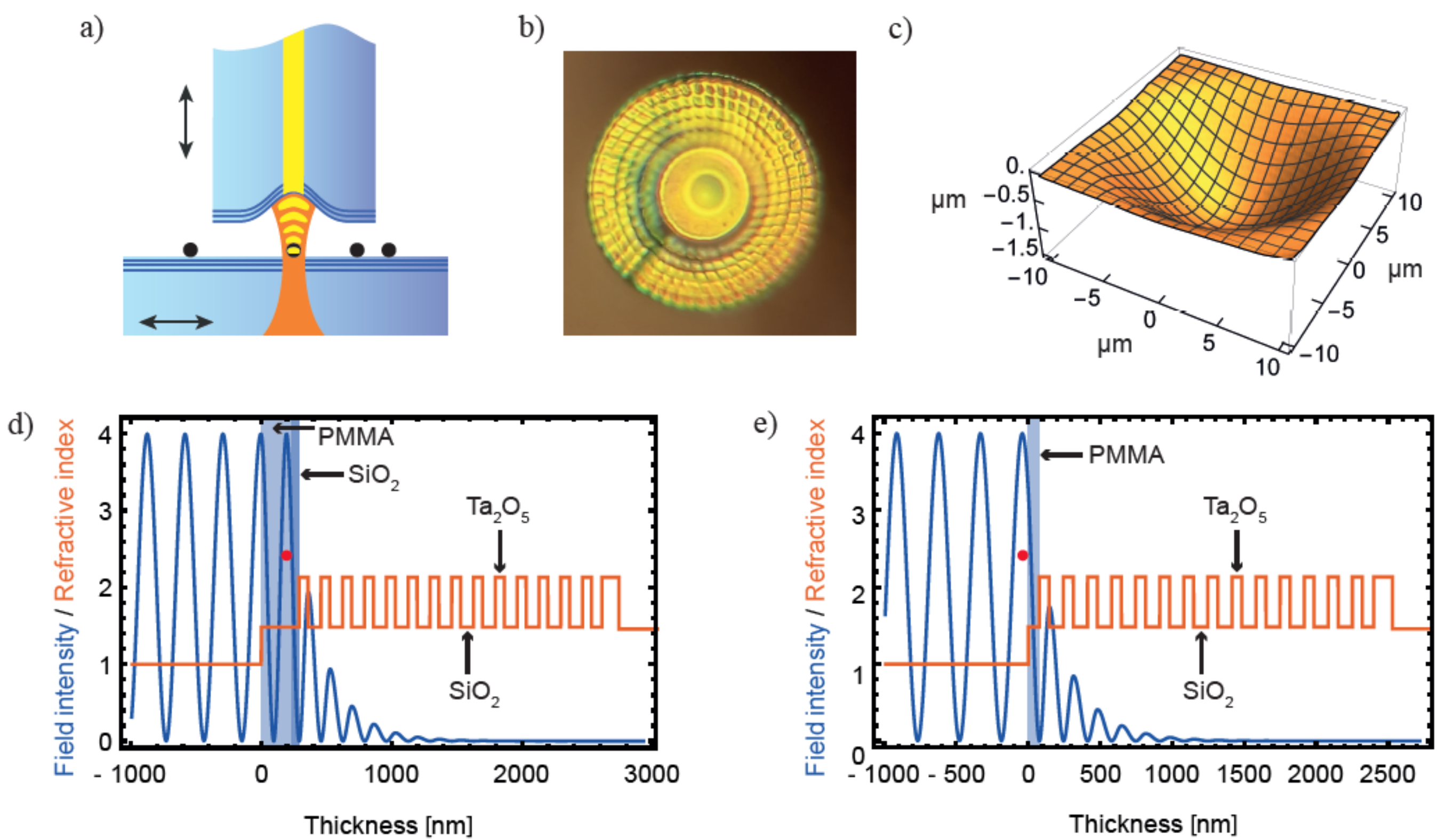}
	\caption{
		{ 
			a) Schematic drawing of the tunable fiber-based microcavity. The mirror can be moved with a three-axis nanopositioning stage, allowing one to set the length of the cavity and locate a crystal inside the cavity.
			b) Optical micrograph of the machined fiber tip, showing the tapering at the outer part and the concave profile in the center.
			c) Measured surface topography of the concave profile. A 2D parabolic fit to the central part yields a radius of curvature of $20~\mu$m.
			d) Refractive index profile (orange) of the mirror and electric field distribution (blue) for an embedded nanocrystal. The red dot marks the location and size of the nanocrystal. 
			e) Same as d) for a nanocrystal on top of a spacer layer as used in the low temperature cavity experiment.
		}
		\label{fig2}}
\end{figure*}

Furthermore, the presence of the nanocrystal and the surrounding dielectric sensitively influence the excited state lifetime of atomic dipoles inside, and thereby affect the light-matter coupling strength. For a nanocrystal far in the Rayleigh regime (diameter $\lesssim20$~nm) embedded in a medium, the lifetime is given by \cite{chew1988,Kien2000,schniepp2002,Duan05}
\be
\tau(x)=\tau_0 n_r(x)\left(\frac{n_r(x)^2+2}{3}\right)^2, \label{taux}
\ee
where $\tau_0$ is the lifetime in bulk Y$_2$O$_3$ and $n_r(x)=n/(x+(1-x)n_m)$ is the relative refractive index with the effective surrounding refractive index $x+(1-x)n_m$ described by the medium refractive index $n_m$ and the filling factor~$x$~\cite{Duan05}. The latter takes into account partially surrounding media, as is the case e.g. for a nanocrystal on a glass substrate. The resulting prolonged lifetime for small crystals is due to the boundary condition of the crystal-air interface, which makes the crystal act as an off-resonant cavity that suppresses the local field inside. According to equation~\ref{taux}, ions in small  Y$_2$O$_3$ nanocrystals in air should show a lifetime $\tau_\mathrm{air}=\tau(0)=7.0\tau_0$. For larger crystals, calculations \cite{Kien2000,Duan05} and measurements \cite{schniepp2002,Duan05} show that $\tau$ decreases with size, and the presence of a surrounding dielectric further reduces $\tau$.

To estimate the effect of embedding, the calculations referred to above are not adequate for our situation, since we do not embed into a bulk medium but into a thin film on a mirror. We therefore perform finite difference time domain simulations (Lumerical) and analyze the spontaneous emission of a dipole located in a 80nm cube of Y$_2$O$_3$ on a Bragg mirror with embedding, see figure \ref{fig2} (d). We find that the emission rate increases by up to a factor of 3.8 as compared to the dipole in a crystal in free space. This shows that embedding increases the local field strength within the crystal, and it will in the same way increase the coupling strength to the cavity field.

We experimentally study these effects for our sample. Figure~\ref{fig1}(e) shows fluorescence lifetime measurements as function of the count rate for crystals prepared in two different ways.
First, the particles are spin coated onto a fused silica substrate to serve as a reference.
Second, we embed the particles into a PMMA layer on top of a dielectric mirror (see figure~\ref{fig2}(d)). 
To ensure that the nanoparticle resides in a field maximum of the standing wave that will be formed in a cavity, a three-step preparation was used: First, a layer of 55 nm of SiO$_2$ was applied to the mirror by electron beam evaporation. Then, nanocrystals are deposited on this layer and are thus centered on an electric field maximum. Third, a 235 nm thick layer of PMMA is spin-coated to cover the particles.
The total optical thickness of the two layers of comparable refractive index $n_m=\{1.46,1.49\}$ is $(3/4)\lambda/n_m$ for the transition wavelength $\lambda=580~$nm. For this thickness, a simulation of the electric field inside the layer \cite{furman1992} shows that it is as large as in the vacuum part of the cavity, thereby maximizing light-matter coupling (see figure~2(d)).

In the experiment, we observe the expected correlation between lifetime and count rate for both samples, when again assuming the relation between count rate and crystal size as discussed above.
The solid line shown in figure~\ref{fig1}(e) agrees with the measurement and is a fit to a prediction of the size-dependent lifetime of a dipole in the center of a dielectric sphere~\cite{Kien2000},
\begin{equation}
\tau(r,x)=\tau_0 n_r(x) U(r),
\end{equation}
with $U(r)$ as given in equation (28) in \cite{Kien2000}. Assuming a single central dipole reasonably approximates the continuous ion distribution in the nanocrystals as long as $r< \lambda/(2\pi n)\approx 50~$nm. 
We use the filling factor $x$ as a fit parameter and obtain best agreement for $x=0.30$. 
This value is smaller than expected, such that other factors that reduce the lifetime in addition to the presence of the substrate cannot be excluded.

We can also observe the effect of embedding the crystals in a thin film, which leads to a marked reduction of the lifetime in the measurement \cite{khalid2015}.
The right panel of Fig.~\ref{fig1}(e) shows the lifetime for the two cases averaged over all crystals, and we observe an averaged lifetime reduction by a factor of 1.5 for the embedded sample, which features a lifetime of $ \tau_\mathrm{emb}=(1.26 \pm 0.03 ) $~ms as compared to $ \tau_\mathrm{sub}=(1.91\pm 0.06) $~ms for the sample on silica. The observed average lifetime reduction is in agreement with the FDTD simulation when considering that the simulation compares a crystal in free space with an embedded one, while the measurement compares crystals on a substrate with embedded ones. 
If we rely on the model used for the fit of the lifetime data, we can use the ratio $\tau_\mathrm{air}/\tau(0)=2.5$ to calculate the expected lifetime in air for the measured data. With this, we obtain an overall lifetime reduction compared to a crystal in air of $(\tau_\mathrm{air}/\tau(0.3))\times (\tau_\mathrm{sub}/\tau_\mathrm{emb}) = 3.75$, close to the FDTD prediction.

\section{Cavity properties}
The main goal of our work is to achieve cavity enhancement of the fluorescence to obtain efficient access to small ensembles and finally individual ions.
To realize such a setting, we have developed a fully tunable, cryogenic-compatible Fabry-Perot microcavity as schematically shown in Figure~\ref{fig2}(a). 
The setup allows us to locate suitable nanocrystals inside the cavity, and to set the mirror separation to a particular longitudinal mode order to achieve resonance conditions with the ions. 
The microcavity consists of a macroscopic planar mirror and a concave micro mirror at the end facet of a single-mode optical fiber. The micro mirror is fabricated by $ \mathrm{CO_2} $-laser machining \cite{hunger2012}, where
we produce a concave profile with a radius of curvature of $20~\mu$m and depth of $1.5~\mu$m. The fiber tip is furthermore shaped into a conical tip to reduce the tip diameter to $\sim 35~\mu$m, which enables a smallest mirror separation of $d=2~\mu$m, see figure \ref{fig2} (b,c).
The fiber and the planar mirror are then coated to have a transmission $T$ of $100~$ppm at $580$~nm as specified and measured by the coating manufacturer. 
The fiber is mounted on a shear piezoelectric actuator for cavity length stabilization, while the large mirror is mounted on a commercial closed-loop three-axis \mbox{attocube} nanopositioning stage.

The finesse of the cavity is measured to be $\mathcal{F}=17,000$ with a variation of $5,000$ depending on the particular fundamental longitudinal mode, values which are in agreement with simulation and calculations, see below. 
The variation of the finesse originates from mode mixing between the fundamental and higher order modes due to the non-spherical shape of the concave structure \cite{benedikter2015}.
From the finesse measurement and the known mirror transmission, we can deduce the residual scattering and absorption losses $L$ of the mirrors and obtain a value $L=30~$ppm.

When the cavity is tuned on resonance, a Purcell factor of $C = 1000 \zeta$ is expected for a cavity with a length of $2~\mu$m, which for $\zeta=1$ quantifies the enhancement of fluorescence of a particular transition. The coherent \transition can thus be significantly enhanced, however due to its small branching ratio $\zeta_{F_0}\approx$ 1/60, the lifetime change is correspondingly smaller, and a Purcell factor $C = 17$ is expected.
In the experiments described here, we were not yet able to stabilize the cavity on resonance in the cryostat, and we investigate an alternative approach, which is less sensitive to the cavity resonance condition. We use the \transition transition for excitation and the \transitionstrong transition for emission, both being simultaneously resonant with the cavity. More detail is given below. This scheme provides high scattering rates also without optimal Purcell enhancement, and reduced background due to the wavelength separation of excitation and detection.

\section{Cavity-enhanced spectroscopy}
To perform low-temperature spectroscopy, we introduce a mirror carrying nanocrystals in the cavity and cool it down to 8~K.
Since the experiments were performed in parallel with the nanocrystal studies described above, an earlier version of sample preparation was used. 
Here, the particles are not embedded within PMMA but reside on top of a PMMA spacer layer, whose thickness is chosen to locate the nanocrystals in an electric field maximum of the cavity modes, see Figure~\ref{fig2}(e). 
To spatially locate a well-isolated crystal of suitable size within the cavity mode, we measure the scattering loss signal of nanocrystals by scanning cavity microscopy~\cite{mader2015}. We evaluate the cavity transmission $T_c$ at $580$~nm, which is given by
\begin{equation}
T_c = \frac{4T_1T_2}{(T_1+T_2 + 2L + 2B)^2}
\end{equation}
where $B$ accounts for the additional scattering loss introduced by the crystal, $T_1=100~$ppm is the fiber mirror transmission, $T_2=190~$ppm the planar mirror transmission, which is increased due to the spacer layer, and $L=30~$ppm the average mirror loss. We calculate $T_2$ from a transfer matrix model \cite{furman1992} that also reproduces $T_1$, and find $L$ from the empty cavity finesse measurement. From the uncertainty of the finesse measurement we find an uncertainty for $T_i,L$ of $<10\%$, and a similar value for $B$.
From the scattering loss $B$ (Equation 1) and the mode waist $w_0$ we calculate the scattering cross section $\sigma$ and the volume of the crystal $V=4/3\pi r^3$ with radius $r$.

\begin{figure*}
\includegraphics[width=0.9\textwidth]{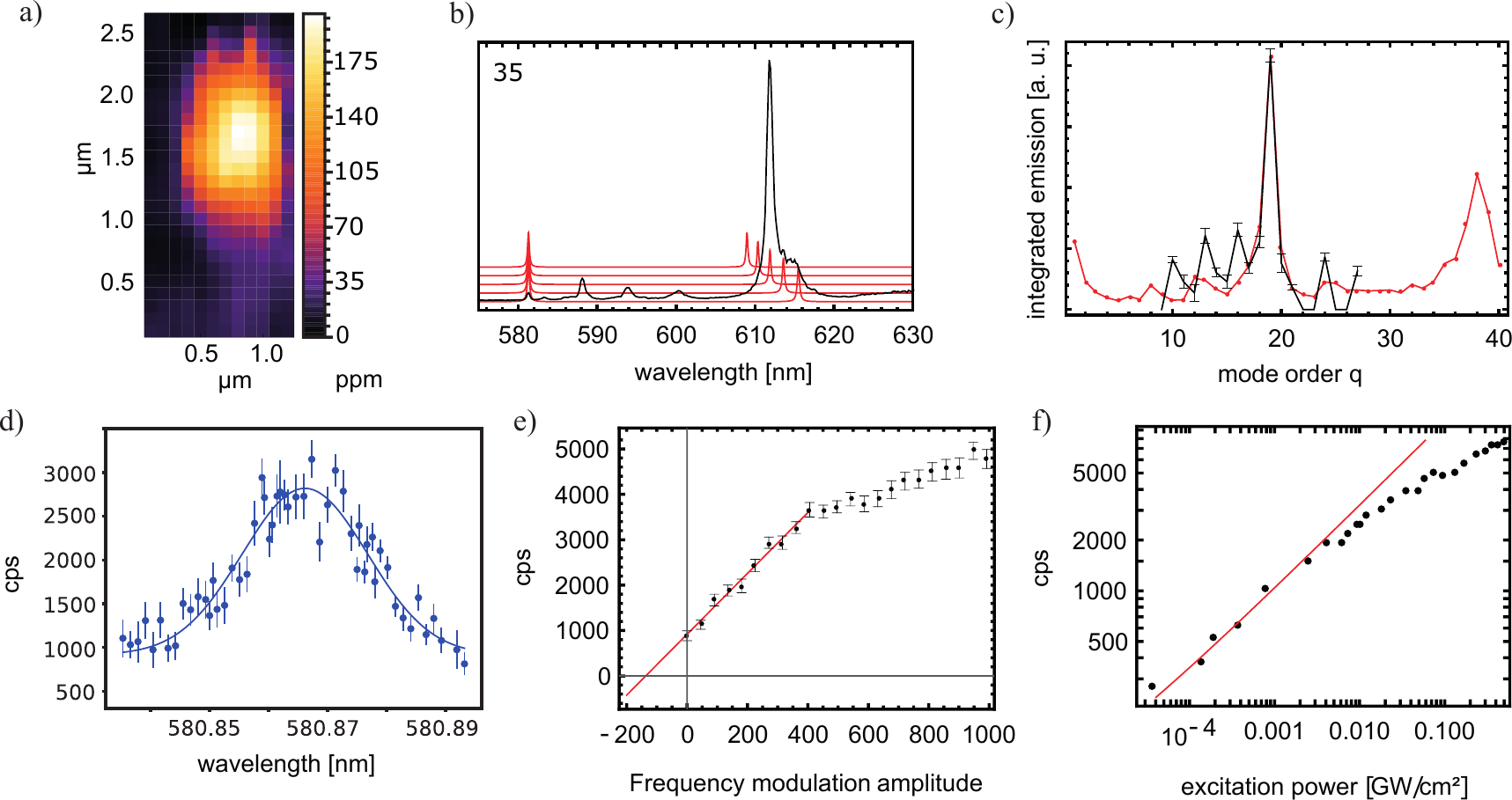}
\caption{
a) Map of scattering loss introduced by a single crystal when scanning the cavity mode across it. b) \eu spectrum (black) overlapped with the cavity spectrum (red) with resonances at $580$~nm and around $611$~nm for longitudinal cavity mode orders $q=17, \dots ,21$ (resonance linewidth not to scale).
c) Calculation (red line) and measurement (black) of the count rate as a function of mode order $q$.
d) Measurement of the inhomogeneous line of the \transition transition. The solid line is a Gaussian fit, yielding a full-width-half-maximum linewidth of 22 GHz.
e) Calibration of the power broadening by laser frequency modulation. The count rate as a function of modulation amplitude (black datapoints) shows a linear dependence (red line: fit). From the extrapolation to zero count rate, we can estimate the power broadening, yielding $140$~MHz in this case.
e) Count rate as function of the intracavity power. 
The solid line is a fit to the data for low power. 
\label{fig3}}
\end{figure*}

We select crystals that are large enough to provide a suitable ion ensemble for fluorescence measurements, but small enough to maintain the out-coupling efficiency of the fluorescence light from the cavity 
\begin{equation}
\eta = \frac{T_2}{ T_1 + T_2 + 2L + 2B}
\end{equation}
at a high level. Figure~\ref{fig3}(a) shows the scattering loss of a single crystal when scanning the cavity mode across it at a cavity length $d=5.5~\mu$m. The shape represents the cavity mode, while the asymmetry is due to a miscalibration of the nanopositioning stage at short length scales.
From the peak loss $B=203\pm20$~ppm observed for perfect spatial alignment, for which $\eta=0.26$, we can calculate the crystal size and obtain a value of $91\pm3$~nm. 
Together with the doping concentration of $0.5\%$, this allows us to estimate the number of ions contained in the crystal, yielding $N= 3\times10^4$ ions in this case.

To obtain a fluorescence signal from the cavity, we take advantage of the dominant emission of the \transitionstrong transition at $611$~nm, which provides larger signals also for an unstabilized cavity. 
We excite the \transition transition at $580$~nm and couple the emission of the \transitionstrong transition to the cavity. 
This detection scheme requires having the cavity resonant with both the excitation light at $580$~nm and the emission light at $611$~nm. 
Figure~\ref{fig3}(b) shows the principle of the scheme, where the \eu spectrum is overlapped with the cavity resonances at $580$~nm and $611$~nm, where at the former wavelength, the longitudinal cavity mode order is varied from $q=17, \dots ,21$. 
For $q=19$, both $580$~nm and $611$~nm light are resonant. 
Figure~\ref{fig3}(c) shows a calculation and a measurement of the count rate as a function of mode order $q$ for $580$~nm cavity excitation.
To perform the measurement, the cavity length was tuned to a particular mode order $q$, and modulated around the set value with an amplitude close to 4nm, which is more than 250 cavity linewidths $\kappa=\frac{c}{2 (q\lambda/2) } \frac{1}{\mathcal F}  \approx 1.3$~GHz at $580$~nm, with a rate of $10$~Hz. We use a single photon counter to collect the fluorescence light from the cavity, and we perform time resolved counting synchronized with the modulation. We make a histogram of the detected counts with a bin size corresponding to about $2\kappa$, and evaluate the averaged peak count rate per bin after several seconds of data acquisition.

Next, we perform spectroscopy of the \transition transition. We set the cavity length to $q=19$ and measure the count rate as describe above as a function of the laser wavelength across the range of the inhomogeneous line of the \transition transition. 
Figure~\ref{fig3}(d) shows the obtained spectrum. 
We fit it with a Gaussian profile and find a FWHM linewidth of 22~GHz, which agrees very well with the linewidth in bulk samples at this doping concentration, and thereby confirms the high crystal quality \cite{kunkel2015}. 
We note that this represents the first low-temperature spectroscopy of a single \eucrystal nanocrystal of such a small size.

To understand the origin of the observed spectrum, we estimate the number of ions in the crystal within the frequency interval of the homogeneous linewidth of a single ion. We assume a Gaussian spectral ion density and neglect the hyperfine structure.
With $N \approx 3 \times 10^4$ ions in the crystal with an expected homogeneous linewidth of about $\gamma_h=200$~kHz at $10$~K \cite{bartholomew2017} spread over an inhomogeneous linewidth of $\gamma_{inh}=22$ GHz.
This estimate leads to a peak density $n=2\sqrt{ln(2)/\pi} N \gamma_h/\gamma_{inh} =0.27$, i.e. the ion transitions are spaced by about four linewidths, and low power spectroscopy would in most cases address no or at most a single ion. 

To obtain the observed signal, we use a very high excitation intensity (several GW/cm${}^2$), which leads to strong power broadening, and thus off resonant excitation of small ion ensembles.
To calibrate the amount of power broadening, we modulate the frequency of the excitation laser and measure the count rate as a function of the modulation amplitude (Figure~\ref{fig3}e). 
We observe a linear signal increase up to an amplitude that approximately corresponds to the cavity linewidth. 
From an extrapolation of the linear dependence to zero count rate, we can estimate the power broadening. 
For the measurement shown, this amounts to $140$~MHz, taken at an intracavity power of $\sim 53$~mW and an intensity of $\sim3.5$~GW/cm${}^2$.

We furthermore measure the count rate on resonance (without frequency modulation) as a function of the excitation power. The scattering rate of a two-level system is given by $R_{sc}=\frac{\gamma}{2}\frac{S_0}{1+S_0+(2\Delta/\gamma)^2}$, and the power broadened linewidth scales with $\sqrt{S_0} =\Omega\sqrt 2/\gamma$, where $S_0$ is the saturation parameter, $\Omega$ the Rabi frequency, and $\gamma$ the excited state decay rate. 
Figure~\ref{fig3}(f) shows a measurement of the count rate as function of the excitation power.
The observed power dependence follows the expected $\sqrt{S_0}$ relation over a large range (red solid line). 
For the weakest accessible excitation power of $0.15$~mW, where we still have a significant signal to background ratio of $2$, the excitation is a factor $400$ weaker than for the value used for the calibration, such that the broadening is a factor $20$ smaller, i.e. $\sim 7$~MHz. With an average spectral ion separation of $\delta = 750$~kHz, this corresponds to probing $\sim 10$~ions. At this power level, we measure $R_\mathrm{det}=300$~counts/s on resonance, such that the single ion count rate is approximately $30$~counts/s. This value is already larger than the estimated single-ion count rate for confocal detection, despite the fact that we collect only a single transition.

We can obtain an order of magnitude estimate of the achieved effective Purcell factor $C$ by comparing the total measured count rate with the expected free space scattering rate after estimating the collection and detection efficiencies and integrating over the power-broadened distribution.
The emission rate in the cavity $R_c$ can be calculated from the inferred total count rate $R_\mathrm{det}$ by 
$R_{c} = R_\mathrm{det}/(\eta \eta_{det})=11000~1/$s,  
where $\eta_{det}=0.1$ includes the detector quantum efficiency, the loss on the optical path, and the cavity outcoupling efficiency $\eta=0.25$. 
Compared to the integrated free space scattering rate for the ensemble under the inferred power broadening, $R_{sc} = \int \gamma/2 (S_0/\delta)1/(1+S_0+(2\Delta/\gamma)^2) d\Delta
 = 5200~1/$s with $\gamma=770$~Hz the measured decay rate and $S_0=93~$MHz, we obtain $C = R_c/R_{sc} \sim  2$.
This shows a net improvement of the signal due to the cavity. 

We can compare the observed value with an estimate of the theoretically achievable Purcell factor for this configuration. At the transition wavelength of $611~$nm, the cavity finesse amounts to $F=5000$, and for the double resonance condition at $q=19$, the mode volume amounts to $35\lambda^3$, such that $C_{F_2}=210 \zeta_{F_2} \approx 80 $ for a branching ratio into the particular crystal field level $\zeta_{F_2}\approx0.4$ as inferred from the spectrum. The linewidth of the \transitionstrong transition is inferred from spectroscopy on a powder and estimated to be $\gamma_{h,F2}>200$~GHz. This is larger than the cavity linewidth $\kappa=1.3~$GHz and reduces the achievable Purcell factor to $C_\mathrm{eff,F_2}=  C_{F_2} \kappa/\gamma_{h,F2}=2.5$, in agreement with the measured value.

\section{Conclusion}
We have performed cavity-enhanced spectroscopy measurements of ensembles of a few  \eu ions in an isolated nanocrystal. 
We characterized the crystal size, such that an estimation of important parameters such as the number of ions probed and the single ion count rate was possible. 
From an order of magnitude estimate, we infer an effective Purcell factor $C \sim 2$, evidencing the benefit of cavity enhancement.
We have also shown that embedding the crystals in a thin film leads to an additional reduction of the lifetime, and calculations show a significant reduction of scattering loss.

A much larger enhancement is expected for an actively stabilized cavity which is continuously coupled to the \transition transition for both excitation and detection.
We are currently developing a custom-built nanopositioning stage with significantly improved passive stability, which has recently enabled us to stabilize the cavity on resonance during the entire cooling cycle of the cryostat, giving good prospects for future experiments. 
For a realistic cavity with $\mathcal F = 10^5$ and $d=2~\mu$m, $C_{F_0}=100$ is expected, such that up to $10^5$ photons/s can be scattered into the cavity, enabling efficient single ion readout. This can open the way to address several individual Eu ions as a multi-qubit register with promising coherence properties.
We are also exploring different materials including \ercrystal, which has a narrow optical transition in the telecom band, ideal for long distance quantum communication applications.

\begin{acknowledgments}
We acknowledge support from T. Bagci and Q. Unterreithmeier during early phases of the experiment. This work was partially funded by the European Union H2020 research and innovation programme under grant agreement No. 712721 (NanOQTech), and the DFG Cluster of Excellence Nanosystems Initiative Munich. 
T.W.H. acknowledges funding from the Max-Planck Foundation, B.C. acknowledges financial support from the Cellex ICFO-MPQ postdoctoral fellowship programme. 
ICFO is also supported by MINECO Severo Ochoa through grant SEV-2015-0522, by fundaci{\'o} Cellex, by CERCA Programme/Generalitat de Catalunya, and from the Spanish Ministerio de Economía y Competitividad (MINECO, FIS2016-81696-ERC).
\end{acknowledgments}


\begin{thebibliography}{35}%
\makeatletter
\providecommand \@ifxundefined [1]{%
 \@ifx{#1\undefined}
}%
\providecommand \@ifnum [1]{%
 \ifnum #1\expandafter \@firstoftwo
 \else \expandafter \@secondoftwo
 \fi
}%
\providecommand \@ifx [1]{%
 \ifx #1\expandafter \@firstoftwo
 \else \expandafter \@secondoftwo
 \fi
}%
\providecommand \natexlab [1]{#1}%
\providecommand \enquote  [1]{``#1''}%
\providecommand \bibnamefont  [1]{#1}%
\providecommand \bibfnamefont [1]{#1}%
\providecommand \citenamefont [1]{#1}%
\providecommand \href@noop [0]{\@secondoftwo}%
\providecommand \href [0]{\begingroup \@sanitize@url \@href}%
\providecommand \@href[1]{\@@startlink{#1}\@@href}%
\providecommand \@@href[1]{\endgroup#1\@@endlink}%
\providecommand \@sanitize@url [0]{\catcode `\\12\catcode `\$12\catcode
  `\&12\catcode `\#12\catcode `\^12\catcode `\_12\catcode `\%12\relax}%
\providecommand \@@startlink[1]{}%
\providecommand \@@endlink[0]{}%
\providecommand \url  [0]{\begingroup\@sanitize@url \@url }%
\providecommand \@url [1]{\endgroup\@href {#1}{\urlprefix }}%
\providecommand \urlprefix  [0]{URL }%
\providecommand \Eprint [0]{\href }%
\providecommand \doibase [0]{http://dx.doi.org/}%
\providecommand \selectlanguage [0]{\@gobble}%
\providecommand \bibinfo  [0]{\@secondoftwo}%
\providecommand \bibfield  [0]{\@secondoftwo}%
\providecommand \translation [1]{[#1]}%
\providecommand \BibitemOpen [0]{}%
\providecommand \bibitemStop [0]{}%
\providecommand \bibitemNoStop [0]{.\EOS\space}%
\providecommand \EOS [0]{\spacefactor3000\relax}%
\providecommand \BibitemShut  [1]{\csname bibitem#1\endcsname}%
\let\auto@bib@innerbib\@empty
\bibitem [{\citenamefont {Goldner}\ \emph {et~al.}(2015)\citenamefont
  {Goldner}, \citenamefont {Ferrier},\ and\ \citenamefont
  {Guillot-Noël}}]{Goldner20151}%
  \BibitemOpen
  \bibfield  {author} {\bibinfo {author} {\bibfnamefont {P.}~\bibnamefont
  {Goldner}}, \bibinfo {author} {\bibfnamefont {A.}~\bibnamefont {Ferrier}}, \
  and\ \bibinfo {author} {\bibfnamefont {O.}~\bibnamefont {Guillot-Noël}},\
  }\href {\doibase https://doi.org/10.1016/B978-0-444-63260-9.00267-4} {}edited
  by\ \bibinfo {editor} {\bibfnamefont {J.-C.~G.}\ \bibnamefont {Bünzli}}\
  and\ \bibinfo {editor} {\bibfnamefont {V.~K.}\ \bibnamefont {Pecharsky}},\
  \bibinfo {series} {Chapter 267 - Rare Earth-Doped Crystals for Quantum
  Information Processing, Handbook on the Physics and Chemistry of Rare
  Earths}, Vol.~\bibinfo {volume} {46}\ (\bibinfo  {publisher} {Elsevier},\
  \bibinfo {year} {2015})\ pp.\ \bibinfo {pages} {1 -- 78}\BibitemShut
  {NoStop}%
\bibitem [{\citenamefont {Zhong}\ \emph {et~al.}(2015)\citenamefont {Zhong},
  \citenamefont {Hedges}, \citenamefont {Ahlefeldt}, \citenamefont
  {Bartholomew}, \citenamefont {Beavan}, \citenamefont {Wittig}, \citenamefont
  {Longdell},\ and\ \citenamefont {Sellars}}]{zhong2015}%
  \BibitemOpen
  \bibfield  {author} {\bibinfo {author} {\bibfnamefont {M.}~\bibnamefont
  {Zhong}}, \bibinfo {author} {\bibfnamefont {M.~P.}\ \bibnamefont {Hedges}},
  \bibinfo {author} {\bibfnamefont {R.~L.}\ \bibnamefont {Ahlefeldt}}, \bibinfo
  {author} {\bibfnamefont {J.~G.}\ \bibnamefont {Bartholomew}}, \bibinfo
  {author} {\bibfnamefont {S.~E.}\ \bibnamefont {Beavan}}, \bibinfo {author}
  {\bibfnamefont {S.~M.}\ \bibnamefont {Wittig}}, \bibinfo {author}
  {\bibfnamefont {J.~J.}\ \bibnamefont {Longdell}}, \ and\ \bibinfo {author}
  {\bibfnamefont {M.~J.}\ \bibnamefont {Sellars}},\ }\href@noop {} {\bibfield
  {journal} {\bibinfo  {journal} {Nature}\ }\textbf {\bibinfo {volume} {517}},\
  \bibinfo {pages} {177} (\bibinfo {year} {2015})}\BibitemShut {NoStop}%
\bibitem [{\citenamefont {Kolesov}\ \emph {et~al.}(2013)\citenamefont
  {Kolesov}, \citenamefont {Xia}, \citenamefont {Reuter}, \citenamefont
  {Jamali}, \citenamefont {St{\"o}hr}, \citenamefont {Inal}, \citenamefont
  {Siyushev},\ and\ \citenamefont {Wrachtrup}}]{kolesov2013}%
  \BibitemOpen
  \bibfield  {author} {\bibinfo {author} {\bibfnamefont {R.}~\bibnamefont
  {Kolesov}}, \bibinfo {author} {\bibfnamefont {K.}~\bibnamefont {Xia}},
  \bibinfo {author} {\bibfnamefont {R.}~\bibnamefont {Reuter}}, \bibinfo
  {author} {\bibfnamefont {M.}~\bibnamefont {Jamali}}, \bibinfo {author}
  {\bibfnamefont {R.}~\bibnamefont {St{\"o}hr}}, \bibinfo {author}
  {\bibfnamefont {T.}~\bibnamefont {Inal}}, \bibinfo {author} {\bibfnamefont
  {P.}~\bibnamefont {Siyushev}}, \ and\ \bibinfo {author} {\bibfnamefont
  {J.}~\bibnamefont {Wrachtrup}},\ }\href@noop {} {\bibfield  {journal}
  {\bibinfo  {journal} {Physical review letters}\ }\textbf {\bibinfo {volume}
  {111}},\ \bibinfo {pages} {120502} (\bibinfo {year} {2013})}\BibitemShut
  {NoStop}%
\bibitem [{\citenamefont {Laplane}\ \emph {et~al.}(2015)\citenamefont
  {Laplane}, \citenamefont {Jobez}, \citenamefont {Etesse}, \citenamefont
  {Timoney}, \citenamefont {Gisin},\ and\ \citenamefont
  {Afzelius}}]{laplane2015}%
  \BibitemOpen
  \bibfield  {author} {\bibinfo {author} {\bibfnamefont {C.}~\bibnamefont
  {Laplane}}, \bibinfo {author} {\bibfnamefont {P.}~\bibnamefont {Jobez}},
  \bibinfo {author} {\bibfnamefont {J.}~\bibnamefont {Etesse}}, \bibinfo
  {author} {\bibfnamefont {N.}~\bibnamefont {Timoney}}, \bibinfo {author}
  {\bibfnamefont {N.}~\bibnamefont {Gisin}}, \ and\ \bibinfo {author}
  {\bibfnamefont {M.}~\bibnamefont {Afzelius}},\ }\href@noop {} {\bibfield
  {journal} {\bibinfo  {journal} {New Journal of Physics}\ }\textbf {\bibinfo
  {volume} {18}},\ \bibinfo {pages} {013006} (\bibinfo {year}
  {2015})}\BibitemShut {NoStop}%
\bibitem [{\citenamefont {G{\"u}ndo{\u{g}}an}\ \emph
  {et~al.}(2015)\citenamefont {G{\"u}ndo{\u{g}}an}, \citenamefont {Ledingham},
  \citenamefont {Kutluer}, \citenamefont {Mazzera},\ and\ \citenamefont
  {de~Riedmatten}}]{gundougan2015}%
  \BibitemOpen
  \bibfield  {author} {\bibinfo {author} {\bibfnamefont {M.}~\bibnamefont
  {G{\"u}ndo{\u{g}}an}}, \bibinfo {author} {\bibfnamefont {P.~M.}\ \bibnamefont
  {Ledingham}}, \bibinfo {author} {\bibfnamefont {K.}~\bibnamefont {Kutluer}},
  \bibinfo {author} {\bibfnamefont {M.}~\bibnamefont {Mazzera}}, \ and\
  \bibinfo {author} {\bibfnamefont {H.}~\bibnamefont {de~Riedmatten}},\
  }\href@noop {} {\bibfield  {journal} {\bibinfo  {journal} {Physical review
  letters}\ }\textbf {\bibinfo {volume} {114}},\ \bibinfo {pages} {230501}
  (\bibinfo {year} {2015})}\BibitemShut {NoStop}%
\bibitem [{\citenamefont {B{\"o}ttger}\ \emph {et~al.}(2009)\citenamefont
  {B{\"o}ttger}, \citenamefont {Thiel}, \citenamefont {Cone},\ and\
  \citenamefont {Sun}}]{bottger2009}%
  \BibitemOpen
  \bibfield  {author} {\bibinfo {author} {\bibfnamefont {T.}~\bibnamefont
  {B{\"o}ttger}}, \bibinfo {author} {\bibfnamefont {C.}~\bibnamefont {Thiel}},
  \bibinfo {author} {\bibfnamefont {R.}~\bibnamefont {Cone}}, \ and\ \bibinfo
  {author} {\bibfnamefont {Y.}~\bibnamefont {Sun}},\ }\href@noop {} {\bibfield
  {journal} {\bibinfo  {journal} {Physical Review B}\ }\textbf {\bibinfo
  {volume} {79}},\ \bibinfo {pages} {115104} (\bibinfo {year}
  {2009})}\BibitemShut {NoStop}%
\bibitem [{\citenamefont {Ohlsson}\ \emph {et~al.}(2002)\citenamefont
  {Ohlsson}, \citenamefont {Mohan},\ and\ \citenamefont
  {Kr{\"o}ll}}]{ohlsson2002}%
  \BibitemOpen
  \bibfield  {author} {\bibinfo {author} {\bibfnamefont {N.}~\bibnamefont
  {Ohlsson}}, \bibinfo {author} {\bibfnamefont {R.~K.}\ \bibnamefont {Mohan}},
  \ and\ \bibinfo {author} {\bibfnamefont {S.}~\bibnamefont {Kr{\"o}ll}},\
  }\href@noop {} {\bibfield  {journal} {\bibinfo  {journal} {Optics
  communications}\ }\textbf {\bibinfo {volume} {201}},\ \bibinfo {pages} {71}
  (\bibinfo {year} {2002})}\BibitemShut {NoStop}%
\bibitem [{\citenamefont {Longdell}\ \emph {et~al.}(2004)\citenamefont
  {Longdell}, \citenamefont {Sellars},\ and\ \citenamefont
  {Manson}}]{longdell2004}%
  \BibitemOpen
  \bibfield  {author} {\bibinfo {author} {\bibfnamefont {J.}~\bibnamefont
  {Longdell}}, \bibinfo {author} {\bibfnamefont {M.}~\bibnamefont {Sellars}}, \
  and\ \bibinfo {author} {\bibfnamefont {N.}~\bibnamefont {Manson}},\
  }\href@noop {} {\bibfield  {journal} {\bibinfo  {journal} {Physical review
  letters}\ }\textbf {\bibinfo {volume} {93}},\ \bibinfo {pages} {130503}
  (\bibinfo {year} {2004})}\BibitemShut {NoStop}%
\bibitem [{\citenamefont {Kolesov}\ \emph {et~al.}(2012)\citenamefont
  {Kolesov}, \citenamefont {Xia}, \citenamefont {Reuter}, \citenamefont
  {St{\"o}hr}, \citenamefont {Zappe}, \citenamefont {Meijer}, \citenamefont
  {Hemmer},\ and\ \citenamefont {Wrachtrup}}]{kolesov2012}%
  \BibitemOpen
  \bibfield  {author} {\bibinfo {author} {\bibfnamefont {R.}~\bibnamefont
  {Kolesov}}, \bibinfo {author} {\bibfnamefont {K.}~\bibnamefont {Xia}},
  \bibinfo {author} {\bibfnamefont {R.}~\bibnamefont {Reuter}}, \bibinfo
  {author} {\bibfnamefont {R.}~\bibnamefont {St{\"o}hr}}, \bibinfo {author}
  {\bibfnamefont {A.}~\bibnamefont {Zappe}}, \bibinfo {author} {\bibfnamefont
  {J.}~\bibnamefont {Meijer}}, \bibinfo {author} {\bibfnamefont
  {P.}~\bibnamefont {Hemmer}}, \ and\ \bibinfo {author} {\bibfnamefont
  {J.}~\bibnamefont {Wrachtrup}},\ }\href@noop {} {\bibfield  {journal}
  {\bibinfo  {journal} {Nature communications}\ }\textbf {\bibinfo {volume}
  {3}},\ \bibinfo {pages} {1029} (\bibinfo {year} {2012})}\BibitemShut
  {NoStop}%
\bibitem [{\citenamefont {Utikal}\ \emph {et~al.}(2014)\citenamefont {Utikal},
  \citenamefont {Eichhammer}, \citenamefont {Petersen}, \citenamefont {Renn},
  \citenamefont {G{\"o}tzinger},\ and\ \citenamefont
  {Sandoghdar}}]{utikal2014}%
  \BibitemOpen
  \bibfield  {author} {\bibinfo {author} {\bibfnamefont {T.}~\bibnamefont
  {Utikal}}, \bibinfo {author} {\bibfnamefont {E.}~\bibnamefont {Eichhammer}},
  \bibinfo {author} {\bibfnamefont {L.}~\bibnamefont {Petersen}}, \bibinfo
  {author} {\bibfnamefont {A.}~\bibnamefont {Renn}}, \bibinfo {author}
  {\bibfnamefont {S.}~\bibnamefont {G{\"o}tzinger}}, \ and\ \bibinfo {author}
  {\bibfnamefont {V.}~\bibnamefont {Sandoghdar}},\ }\href@noop {} {\bibfield
  {journal} {\bibinfo  {journal} {Nature communications}\ }\textbf {\bibinfo
  {volume} {5}},\ \bibinfo {pages} {3627} (\bibinfo {year} {2014})}\BibitemShut
  {NoStop}%
\bibitem [{\citenamefont {Nakamura}\ \emph {et~al.}(2014)\citenamefont
  {Nakamura}, \citenamefont {Yoshihiro}, \citenamefont {Inagawa}, \citenamefont
  {Fujiyoshi},\ and\ \citenamefont {Matsushita}}]{nakamura2014}%
  \BibitemOpen
  \bibfield  {author} {\bibinfo {author} {\bibfnamefont {I.}~\bibnamefont
  {Nakamura}}, \bibinfo {author} {\bibfnamefont {T.}~\bibnamefont {Yoshihiro}},
  \bibinfo {author} {\bibfnamefont {H.}~\bibnamefont {Inagawa}}, \bibinfo
  {author} {\bibfnamefont {S.}~\bibnamefont {Fujiyoshi}}, \ and\ \bibinfo
  {author} {\bibfnamefont {M.}~\bibnamefont {Matsushita}},\ }\href@noop {}
  {\bibfield  {journal} {\bibinfo  {journal} {Scientific reports}\ }\textbf
  {\bibinfo {volume} {4}},\ \bibinfo {pages} {7364} (\bibinfo {year}
  {2014})}\BibitemShut {NoStop}%
\bibitem [{\citenamefont {Siyushev}\ \emph {et~al.}(2014)\citenamefont
  {Siyushev}, \citenamefont {Xia}, \citenamefont {Reuter}, \citenamefont
  {Jamali}, \citenamefont {Zhao}, \citenamefont {Yang}, \citenamefont {Duan},
  \citenamefont {Kukharchyk}, \citenamefont {Wieck}, \citenamefont {Kolesov}
  \emph {et~al.}}]{siyushev2014}%
  \BibitemOpen
  \bibfield  {author} {\bibinfo {author} {\bibfnamefont {P.}~\bibnamefont
  {Siyushev}}, \bibinfo {author} {\bibfnamefont {K.}~\bibnamefont {Xia}},
  \bibinfo {author} {\bibfnamefont {R.}~\bibnamefont {Reuter}}, \bibinfo
  {author} {\bibfnamefont {M.}~\bibnamefont {Jamali}}, \bibinfo {author}
  {\bibfnamefont {N.}~\bibnamefont {Zhao}}, \bibinfo {author} {\bibfnamefont
  {N.}~\bibnamefont {Yang}}, \bibinfo {author} {\bibfnamefont {C.}~\bibnamefont
  {Duan}}, \bibinfo {author} {\bibfnamefont {N.}~\bibnamefont {Kukharchyk}},
  \bibinfo {author} {\bibfnamefont {A.}~\bibnamefont {Wieck}}, \bibinfo
  {author} {\bibfnamefont {R.}~\bibnamefont {Kolesov}},  \emph {et~al.},\
  }\href@noop {} {\bibfield  {journal} {\bibinfo  {journal} {Nature
  communications}\ }\textbf {\bibinfo {volume} {5}},\ \bibinfo {pages} {3895}
  (\bibinfo {year} {2014})}\BibitemShut {NoStop}%
\bibitem [{\citenamefont {Dibos}\ \emph {et~al.}(2017)\citenamefont {Dibos},
  \citenamefont {Raha}, \citenamefont {Phenicie},\ and\ \citenamefont
  {Thompson}}]{dibos2017}%
  \BibitemOpen
  \bibfield  {author} {\bibinfo {author} {\bibfnamefont {A.}~\bibnamefont
  {Dibos}}, \bibinfo {author} {\bibfnamefont {M.}~\bibnamefont {Raha}},
  \bibinfo {author} {\bibfnamefont {C.}~\bibnamefont {Phenicie}}, \ and\
  \bibinfo {author} {\bibfnamefont {J.}~\bibnamefont {Thompson}},\ }\href@noop
  {} {\bibfield  {journal} {\bibinfo  {journal} {arXiv preprint
  arXiv:1711.10368}\ } (\bibinfo {year} {2017})}\BibitemShut {NoStop}%
\bibitem [{\citenamefont {Zhong}\ \emph {et~al.}(2018)\citenamefont {Zhong},
  \citenamefont {Kindem}, \citenamefont {Bartholomew}, \citenamefont {Rochman},
  \citenamefont {Craiciu}, \citenamefont {Verma}, \citenamefont {Nam},
  \citenamefont {Marsili}, \citenamefont {Shaw}, \citenamefont {Beyer} \emph
  {et~al.}}]{zhong2018}%
  \BibitemOpen
  \bibfield  {author} {\bibinfo {author} {\bibfnamefont {T.}~\bibnamefont
  {Zhong}}, \bibinfo {author} {\bibfnamefont {J.~M.}\ \bibnamefont {Kindem}},
  \bibinfo {author} {\bibfnamefont {J.~G.}\ \bibnamefont {Bartholomew}},
  \bibinfo {author} {\bibfnamefont {J.}~\bibnamefont {Rochman}}, \bibinfo
  {author} {\bibfnamefont {I.}~\bibnamefont {Craiciu}}, \bibinfo {author}
  {\bibfnamefont {V.}~\bibnamefont {Verma}}, \bibinfo {author} {\bibfnamefont
  {S.~W.}\ \bibnamefont {Nam}}, \bibinfo {author} {\bibfnamefont
  {F.}~\bibnamefont {Marsili}}, \bibinfo {author} {\bibfnamefont {M.~D.}\
  \bibnamefont {Shaw}}, \bibinfo {author} {\bibfnamefont {A.~D.}\ \bibnamefont
  {Beyer}},  \emph {et~al.},\ }\href@noop {} {\bibfield  {journal} {\bibinfo
  {journal} {arXiv preprint arXiv:1803.07520}\ } (\bibinfo {year}
  {2018})}\BibitemShut {NoStop}%
\bibitem [{\citenamefont {Zukowski}\ \emph {et~al.}(1993)\citenamefont
  {Zukowski}, \citenamefont {Zeilinger}, \citenamefont {Horne},\ and\
  \citenamefont {Ekert}}]{zukowski1993}%
  \BibitemOpen
  \bibfield  {author} {\bibinfo {author} {\bibfnamefont {M.}~\bibnamefont
  {Zukowski}}, \bibinfo {author} {\bibfnamefont {A.}~\bibnamefont {Zeilinger}},
  \bibinfo {author} {\bibfnamefont {M.~A.}\ \bibnamefont {Horne}}, \ and\
  \bibinfo {author} {\bibfnamefont {A.~K.}\ \bibnamefont {Ekert}},\ }\href@noop
  {} {\bibfield  {journal} {\bibinfo  {journal} {Physical Review Letters}\
  }\textbf {\bibinfo {volume} {71}},\ \bibinfo {pages} {4287} (\bibinfo {year}
  {1993})}\BibitemShut {NoStop}%
\bibitem [{\citenamefont {Pan}\ \emph {et~al.}(1998)\citenamefont {Pan},
  \citenamefont {Bouwmeester}, \citenamefont {Weinfurter},\ and\ \citenamefont
  {Zeilinger}}]{pan1998}%
  \BibitemOpen
  \bibfield  {author} {\bibinfo {author} {\bibfnamefont {J.-W.}\ \bibnamefont
  {Pan}}, \bibinfo {author} {\bibfnamefont {D.}~\bibnamefont {Bouwmeester}},
  \bibinfo {author} {\bibfnamefont {H.}~\bibnamefont {Weinfurter}}, \ and\
  \bibinfo {author} {\bibfnamefont {A.}~\bibnamefont {Zeilinger}},\ }\href@noop
  {} {\bibfield  {journal} {\bibinfo  {journal} {Physical Review Letters}\
  }\textbf {\bibinfo {volume} {80}},\ \bibinfo {pages} {3891} (\bibinfo {year}
  {1998})}\BibitemShut {NoStop}%
\bibitem [{\citenamefont {Olmschenk}\ \emph {et~al.}(2009)\citenamefont
  {Olmschenk}, \citenamefont {Matsukevich}, \citenamefont {Maunz},
  \citenamefont {Hayes}, \citenamefont {Duan},\ and\ \citenamefont
  {Monroe}}]{olmschenk2009}%
  \BibitemOpen
  \bibfield  {author} {\bibinfo {author} {\bibfnamefont {S.}~\bibnamefont
  {Olmschenk}}, \bibinfo {author} {\bibfnamefont {D.}~\bibnamefont
  {Matsukevich}}, \bibinfo {author} {\bibfnamefont {P.}~\bibnamefont {Maunz}},
  \bibinfo {author} {\bibfnamefont {D.}~\bibnamefont {Hayes}}, \bibinfo
  {author} {\bibfnamefont {L.-M.}\ \bibnamefont {Duan}}, \ and\ \bibinfo
  {author} {\bibfnamefont {C.}~\bibnamefont {Monroe}},\ }\href@noop {}
  {\bibfield  {journal} {\bibinfo  {journal} {Science}\ }\textbf {\bibinfo
  {volume} {323}},\ \bibinfo {pages} {486} (\bibinfo {year}
  {2009})}\BibitemShut {NoStop}%
\bibitem [{\citenamefont {Hunger}\ \emph {et~al.}(2010)\citenamefont {Hunger},
  \citenamefont {Steinmetz}, \citenamefont {Colombe}, \citenamefont {Deutsch},
  \citenamefont {H{\"a}nsch},\ and\ \citenamefont {Reichel}}]{hunger2010}%
  \BibitemOpen
  \bibfield  {author} {\bibinfo {author} {\bibfnamefont {D.}~\bibnamefont
  {Hunger}}, \bibinfo {author} {\bibfnamefont {T.}~\bibnamefont {Steinmetz}},
  \bibinfo {author} {\bibfnamefont {Y.}~\bibnamefont {Colombe}}, \bibinfo
  {author} {\bibfnamefont {C.}~\bibnamefont {Deutsch}}, \bibinfo {author}
  {\bibfnamefont {T.~W.}\ \bibnamefont {H{\"a}nsch}}, \ and\ \bibinfo {author}
  {\bibfnamefont {J.}~\bibnamefont {Reichel}},\ }\href@noop {} {\bibfield
  {journal} {\bibinfo  {journal} {New Journal of Physics}\ }\textbf {\bibinfo
  {volume} {12}},\ \bibinfo {pages} {065038} (\bibinfo {year}
  {2010})}\BibitemShut {NoStop}%
\bibitem [{\citenamefont {Benedikter}\ \emph {et~al.}(2017)\citenamefont
  {Benedikter}, \citenamefont {Kaupp}, \citenamefont {H{\"u}mmer},
  \citenamefont {Liang}, \citenamefont {Bommer}, \citenamefont {Becher},
  \citenamefont {Krueger}, \citenamefont {Smith}, \citenamefont {H{\"a}nsch},\
  and\ \citenamefont {Hunger}}]{benedikter2017}%
  \BibitemOpen
  \bibfield  {author} {\bibinfo {author} {\bibfnamefont {J.}~\bibnamefont
  {Benedikter}}, \bibinfo {author} {\bibfnamefont {H.}~\bibnamefont {Kaupp}},
  \bibinfo {author} {\bibfnamefont {T.}~\bibnamefont {H{\"u}mmer}}, \bibinfo
  {author} {\bibfnamefont {Y.}~\bibnamefont {Liang}}, \bibinfo {author}
  {\bibfnamefont {A.}~\bibnamefont {Bommer}}, \bibinfo {author} {\bibfnamefont
  {C.}~\bibnamefont {Becher}}, \bibinfo {author} {\bibfnamefont
  {A.}~\bibnamefont {Krueger}}, \bibinfo {author} {\bibfnamefont {J.~M.}\
  \bibnamefont {Smith}}, \bibinfo {author} {\bibfnamefont {T.~W.}\ \bibnamefont
  {H{\"a}nsch}}, \ and\ \bibinfo {author} {\bibfnamefont {D.}~\bibnamefont
  {Hunger}},\ }\href@noop {} {\bibfield  {journal} {\bibinfo  {journal}
  {Physical Review Applied}\ }\textbf {\bibinfo {volume} {7}},\ \bibinfo
  {pages} {024031} (\bibinfo {year} {2017})}\BibitemShut {NoStop}%
\bibitem [{\citenamefont {Perrot}\ \emph {et~al.}(2013)\citenamefont {Perrot},
  \citenamefont {Goldner}, \citenamefont {Giaume}, \citenamefont {Lovri{\'c}},
  \citenamefont {Andriamiadamanana}, \citenamefont {Gon{\c{c}}alves},\ and\
  \citenamefont {Ferrier}}]{perrot2013}%
  \BibitemOpen
  \bibfield  {author} {\bibinfo {author} {\bibfnamefont {A.}~\bibnamefont
  {Perrot}}, \bibinfo {author} {\bibfnamefont {P.}~\bibnamefont {Goldner}},
  \bibinfo {author} {\bibfnamefont {D.}~\bibnamefont {Giaume}}, \bibinfo
  {author} {\bibfnamefont {M.}~\bibnamefont {Lovri{\'c}}}, \bibinfo {author}
  {\bibfnamefont {C.}~\bibnamefont {Andriamiadamanana}}, \bibinfo {author}
  {\bibfnamefont {R.}~\bibnamefont {Gon{\c{c}}alves}}, \ and\ \bibinfo {author}
  {\bibfnamefont {A.}~\bibnamefont {Ferrier}},\ }\href@noop {} {\bibfield
  {journal} {\bibinfo  {journal} {Physical review letters}\ }\textbf {\bibinfo
  {volume} {111}},\ \bibinfo {pages} {203601} (\bibinfo {year}
  {2013})}\BibitemShut {NoStop}%
\bibitem [{\citenamefont {Bartholomew}\ \emph {et~al.}(2017)\citenamefont
  {Bartholomew}, \citenamefont {de~Oliveira~Lima}, \citenamefont {Ferrier},\
  and\ \citenamefont {Goldner}}]{bartholomew2017}%
  \BibitemOpen
  \bibfield  {author} {\bibinfo {author} {\bibfnamefont {J.~G.}\ \bibnamefont
  {Bartholomew}}, \bibinfo {author} {\bibfnamefont {K.}~\bibnamefont
  {de~Oliveira~Lima}}, \bibinfo {author} {\bibfnamefont {A.}~\bibnamefont
  {Ferrier}}, \ and\ \bibinfo {author} {\bibfnamefont {P.}~\bibnamefont
  {Goldner}},\ }\href@noop {} {\bibfield  {journal} {\bibinfo  {journal} {Nano
  letters}\ }\textbf {\bibinfo {volume} {17}},\ \bibinfo {pages} {778}
  (\bibinfo {year} {2017})}\BibitemShut {NoStop}%
\bibitem [{\citenamefont {Serrano}\ \emph {et~al.}(2017)\citenamefont
  {Serrano}, \citenamefont {Karlsson}, \citenamefont {Fossati}, \citenamefont
  {Ferrier},\ and\ \citenamefont {Goldner}}]{serrano2017}%
  \BibitemOpen
  \bibfield  {author} {\bibinfo {author} {\bibfnamefont {D.}~\bibnamefont
  {Serrano}}, \bibinfo {author} {\bibfnamefont {J.}~\bibnamefont {Karlsson}},
  \bibinfo {author} {\bibfnamefont {A.}~\bibnamefont {Fossati}}, \bibinfo
  {author} {\bibfnamefont {A.}~\bibnamefont {Ferrier}}, \ and\ \bibinfo
  {author} {\bibfnamefont {P.}~\bibnamefont {Goldner}},\ }\href@noop {}
  {\bibfield  {journal} {\bibinfo  {journal} {arXiv preprint arXiv:1711.03934}\
  } (\bibinfo {year} {2017})}\BibitemShut {NoStop}%
\bibitem [{\citenamefont {Lauritzen}\ \emph {et~al.}(2012)\citenamefont
  {Lauritzen}, \citenamefont {Timoney}, \citenamefont {Gisin}, \citenamefont
  {Afzelius}, \citenamefont {de~Riedmatten}, \citenamefont {Sun}, \citenamefont
  {Macfarlane},\ and\ \citenamefont {Cone}}]{lauritzen2012}%
  \BibitemOpen
  \bibfield  {author} {\bibinfo {author} {\bibfnamefont {B.}~\bibnamefont
  {Lauritzen}}, \bibinfo {author} {\bibfnamefont {N.}~\bibnamefont {Timoney}},
  \bibinfo {author} {\bibfnamefont {N.}~\bibnamefont {Gisin}}, \bibinfo
  {author} {\bibfnamefont {M.}~\bibnamefont {Afzelius}}, \bibinfo {author}
  {\bibfnamefont {H.}~\bibnamefont {de~Riedmatten}}, \bibinfo {author}
  {\bibfnamefont {Y.}~\bibnamefont {Sun}}, \bibinfo {author} {\bibfnamefont
  {R.}~\bibnamefont {Macfarlane}}, \ and\ \bibinfo {author} {\bibfnamefont
  {R.}~\bibnamefont {Cone}},\ }\href@noop {} {\bibfield  {journal} {\bibinfo
  {journal} {Physical Review B}\ }\textbf {\bibinfo {volume} {85}},\ \bibinfo
  {pages} {115111} (\bibinfo {year} {2012})}\BibitemShut {NoStop}%
\bibitem [{\citenamefont {de~Oliveira~Lima}\ \emph {et~al.}(2015)\citenamefont
  {de~Oliveira~Lima}, \citenamefont {Gon{\c{c}}alves}, \citenamefont {Giaume},
  \citenamefont {Ferrier},\ and\ \citenamefont {Goldner}}]{de2015}%
  \BibitemOpen
  \bibfield  {author} {\bibinfo {author} {\bibfnamefont {K.}~\bibnamefont
  {de~Oliveira~Lima}}, \bibinfo {author} {\bibfnamefont {R.~R.}\ \bibnamefont
  {Gon{\c{c}}alves}}, \bibinfo {author} {\bibfnamefont {D.}~\bibnamefont
  {Giaume}}, \bibinfo {author} {\bibfnamefont {A.}~\bibnamefont {Ferrier}}, \
  and\ \bibinfo {author} {\bibfnamefont {P.}~\bibnamefont {Goldner}},\
  }\href@noop {} {\bibfield  {journal} {\bibinfo  {journal} {Journal of
  Luminescence}\ }\textbf {\bibinfo {volume} {168}},\ \bibinfo {pages} {276}
  (\bibinfo {year} {2015})}\BibitemShut {NoStop}%
\bibitem [{\citenamefont {Wind}\ \emph {et~al.}(1987)\citenamefont {Wind},
  \citenamefont {Vlieger},\ and\ \citenamefont {Bedeaux}}]{wind1987}%
  \BibitemOpen
  \bibfield  {author} {\bibinfo {author} {\bibfnamefont {M.}~\bibnamefont
  {Wind}}, \bibinfo {author} {\bibfnamefont {J.}~\bibnamefont {Vlieger}}, \
  and\ \bibinfo {author} {\bibfnamefont {D.}~\bibnamefont {Bedeaux}},\
  }\href@noop {} {\bibfield  {journal} {\bibinfo  {journal} {Physica A:
  Statistical Mechanics and its Applications}\ }\textbf {\bibinfo {volume}
  {141}},\ \bibinfo {pages} {33} (\bibinfo {year} {1987})}\BibitemShut
  {NoStop}%
\bibitem [{\citenamefont {Chew}(1988)}]{chew1988}%
  \BibitemOpen
  \bibfield  {author} {\bibinfo {author} {\bibfnamefont {H.}~\bibnamefont
  {Chew}},\ }\href@noop {} {\bibfield  {journal} {\bibinfo  {journal} {Physical
  Review A}\ }\textbf {\bibinfo {volume} {38}},\ \bibinfo {pages} {3410}
  (\bibinfo {year} {1988})}\BibitemShut {NoStop}%
\bibitem [{\citenamefont {Le~Kien}\ \emph {et~al.}(2000)\citenamefont
  {Le~Kien}, \citenamefont {Quang},\ and\ \citenamefont {Hakuta}}]{Kien2000}%
  \BibitemOpen
  \bibfield  {author} {\bibinfo {author} {\bibfnamefont {F.}~\bibnamefont
  {Le~Kien}}, \bibinfo {author} {\bibfnamefont {N.~H.}\ \bibnamefont {Quang}},
  \ and\ \bibinfo {author} {\bibfnamefont {K.}~\bibnamefont {Hakuta}},\
  }\href@noop {} {\bibfield  {journal} {\bibinfo  {journal} {Optics
  communications}\ }\textbf {\bibinfo {volume} {178}},\ \bibinfo {pages} {151}
  (\bibinfo {year} {2000})}\BibitemShut {NoStop}%
\bibitem [{\citenamefont {Schniepp}\ and\ \citenamefont
  {Sandoghdar}(2002)}]{schniepp2002}%
  \BibitemOpen
  \bibfield  {author} {\bibinfo {author} {\bibfnamefont {H.}~\bibnamefont
  {Schniepp}}\ and\ \bibinfo {author} {\bibfnamefont {V.}~\bibnamefont
  {Sandoghdar}},\ }\href@noop {} {\bibfield  {journal} {\bibinfo  {journal}
  {Physical review letters}\ }\textbf {\bibinfo {volume} {89}},\ \bibinfo
  {pages} {257403} (\bibinfo {year} {2002})}\BibitemShut {NoStop}%
\bibitem [{\citenamefont {Duan}\ \emph {et~al.}(2005)\citenamefont {Duan},
  \citenamefont {Reid},\ and\ \citenamefont {Wang}}]{Duan05}%
  \BibitemOpen
  \bibfield  {author} {\bibinfo {author} {\bibfnamefont {C.-K.}\ \bibnamefont
  {Duan}}, \bibinfo {author} {\bibfnamefont {M.~F.}\ \bibnamefont {Reid}}, \
  and\ \bibinfo {author} {\bibfnamefont {Z.}~\bibnamefont {Wang}},\ }\href@noop
  {} {\bibfield  {journal} {\bibinfo  {journal} {Physics Letters A}\ }\textbf
  {\bibinfo {volume} {343}},\ \bibinfo {pages} {474} (\bibinfo {year}
  {2005})}\BibitemShut {NoStop}%
\bibitem [{\citenamefont {Furman}\ and\ \citenamefont
  {Tikhonravov}(1992)}]{furman1992}%
  \BibitemOpen
  \bibfield  {author} {\bibinfo {author} {\bibfnamefont {S.~A.}\ \bibnamefont
  {Furman}}\ and\ \bibinfo {author} {\bibfnamefont {A.}~\bibnamefont
  {Tikhonravov}},\ }\href@noop {} {\emph {\bibinfo {title} {Basics of optics of
  multilayer systems}}}\ (\bibinfo  {publisher} {Atlantica S{\'e}guier
  Frontieres},\ \bibinfo {year} {1992})\BibitemShut {NoStop}%
\bibitem [{\citenamefont {Khalid}\ \emph {et~al.}(2015)\citenamefont {Khalid},
  \citenamefont {Chung}, \citenamefont {Rajasekharan}, \citenamefont {Lau},
  \citenamefont {Karle}, \citenamefont {Gibson},\ and\ \citenamefont
  {Tomljenovic-Hanic}}]{khalid2015}%
  \BibitemOpen
  \bibfield  {author} {\bibinfo {author} {\bibfnamefont {A.}~\bibnamefont
  {Khalid}}, \bibinfo {author} {\bibfnamefont {K.}~\bibnamefont {Chung}},
  \bibinfo {author} {\bibfnamefont {R.}~\bibnamefont {Rajasekharan}}, \bibinfo
  {author} {\bibfnamefont {D.~W.}\ \bibnamefont {Lau}}, \bibinfo {author}
  {\bibfnamefont {T.~J.}\ \bibnamefont {Karle}}, \bibinfo {author}
  {\bibfnamefont {B.~C.}\ \bibnamefont {Gibson}}, \ and\ \bibinfo {author}
  {\bibfnamefont {S.}~\bibnamefont {Tomljenovic-Hanic}},\ }\href@noop {}
  {\bibfield  {journal} {\bibinfo  {journal} {Scientific reports}\ }\textbf
  {\bibinfo {volume} {5}},\ \bibinfo {pages} {11179} (\bibinfo {year}
  {2015})}\BibitemShut {NoStop}%
\bibitem [{\citenamefont {Hunger}\ \emph {et~al.}(2012)\citenamefont {Hunger},
  \citenamefont {Deutsch}, \citenamefont {Barbour}, \citenamefont {Warburton},\
  and\ \citenamefont {Reichel}}]{hunger2012}%
  \BibitemOpen
  \bibfield  {author} {\bibinfo {author} {\bibfnamefont {D.}~\bibnamefont
  {Hunger}}, \bibinfo {author} {\bibfnamefont {C.}~\bibnamefont {Deutsch}},
  \bibinfo {author} {\bibfnamefont {R.~J.}\ \bibnamefont {Barbour}}, \bibinfo
  {author} {\bibfnamefont {R.~J.}\ \bibnamefont {Warburton}}, \ and\ \bibinfo
  {author} {\bibfnamefont {J.}~\bibnamefont {Reichel}},\ }\href@noop {}
  {\bibfield  {journal} {\bibinfo  {journal} {Aip Advances}\ }\textbf {\bibinfo
  {volume} {2}},\ \bibinfo {pages} {012119} (\bibinfo {year}
  {2012})}\BibitemShut {NoStop}%
\bibitem [{\citenamefont {Benedikter}\ \emph {et~al.}(2015)\citenamefont
  {Benedikter}, \citenamefont {H{\"u}mmer}, \citenamefont {Mader},
  \citenamefont {Schlederer}, \citenamefont {Reichel}, \citenamefont
  {H{\"a}nsch},\ and\ \citenamefont {Hunger}}]{benedikter2015}%
  \BibitemOpen
  \bibfield  {author} {\bibinfo {author} {\bibfnamefont {J.}~\bibnamefont
  {Benedikter}}, \bibinfo {author} {\bibfnamefont {T.}~\bibnamefont
  {H{\"u}mmer}}, \bibinfo {author} {\bibfnamefont {M.}~\bibnamefont {Mader}},
  \bibinfo {author} {\bibfnamefont {B.}~\bibnamefont {Schlederer}}, \bibinfo
  {author} {\bibfnamefont {J.}~\bibnamefont {Reichel}}, \bibinfo {author}
  {\bibfnamefont {T.~W.}\ \bibnamefont {H{\"a}nsch}}, \ and\ \bibinfo {author}
  {\bibfnamefont {D.}~\bibnamefont {Hunger}},\ }\href@noop {} {\bibfield
  {journal} {\bibinfo  {journal} {New Journal of Physics}\ }\textbf {\bibinfo
  {volume} {17}},\ \bibinfo {pages} {053051} (\bibinfo {year}
  {2015})}\BibitemShut {NoStop}%
\bibitem [{\citenamefont {Mader}\ \emph {et~al.}(2015)\citenamefont {Mader},
  \citenamefont {Reichel}, \citenamefont {H{\"a}nsch},\ and\ \citenamefont
  {Hunger}}]{mader2015}%
  \BibitemOpen
  \bibfield  {author} {\bibinfo {author} {\bibfnamefont {M.}~\bibnamefont
  {Mader}}, \bibinfo {author} {\bibfnamefont {J.}~\bibnamefont {Reichel}},
  \bibinfo {author} {\bibfnamefont {T.~W.}\ \bibnamefont {H{\"a}nsch}}, \ and\
  \bibinfo {author} {\bibfnamefont {D.}~\bibnamefont {Hunger}},\ }\href@noop {}
  {\bibfield  {journal} {\bibinfo  {journal} {Nature communications}\ }\textbf
  {\bibinfo {volume} {6}},\ \bibinfo {pages} {7249} (\bibinfo {year}
  {2015})}\BibitemShut {NoStop}%
\bibitem [{\citenamefont {Kunkel}\ \emph {et~al.}(2015)\citenamefont {Kunkel},
  \citenamefont {Ferrier}, \citenamefont {Thiel}, \citenamefont {Ram{\'\i}rez},
  \citenamefont {Bausa}, \citenamefont {Cone}, \citenamefont {Ikesue},\ and\
  \citenamefont {Goldner}}]{kunkel2015}%
  \BibitemOpen
  \bibfield  {author} {\bibinfo {author} {\bibfnamefont {N.}~\bibnamefont
  {Kunkel}}, \bibinfo {author} {\bibfnamefont {A.}~\bibnamefont {Ferrier}},
  \bibinfo {author} {\bibfnamefont {C.~W.}\ \bibnamefont {Thiel}}, \bibinfo
  {author} {\bibfnamefont {M.~O.}\ \bibnamefont {Ram{\'\i}rez}}, \bibinfo
  {author} {\bibfnamefont {L.~E.}\ \bibnamefont {Bausa}}, \bibinfo {author}
  {\bibfnamefont {R.~L.}\ \bibnamefont {Cone}}, \bibinfo {author}
  {\bibfnamefont {A.}~\bibnamefont {Ikesue}}, \ and\ \bibinfo {author}
  {\bibfnamefont {P.}~\bibnamefont {Goldner}},\ }\href@noop {} {\bibfield
  {journal} {\bibinfo  {journal} {APL Materials}\ }\textbf {\bibinfo {volume}
  {3}},\ \bibinfo {pages} {096103} (\bibinfo {year} {2015})}\BibitemShut
  {NoStop}%
\end{thebibliography}
\end{document}